# Computer modelling of connectivity change suggests epileptogenesis mechanisms in idiopathic generalised epilepsy

Nishant Sinha [1a,b], Yujiang Wang[a,b,c], Justin Dauwels[d], Marcus Kaiser[a,b], Thomas Thesen[e,f], Rob Forsyth[a], Peter Neal Taylor [2a,b,c]

[a]*Institute of Neuroscience, Faculty of Medical Sciences, Newcastle University, Newcastle upon Tyne, UK*
[b]*ICOS, School of Computing, Newcastle University, UK*
[c]*Institute of Neurology, University College London, UK*
[d]*School of Electrical and Electronic Engineering, Nanyang Technological University, Singapore*
[e]*Department of Neurology, School of Medicine, New York University, NY, USA*
[f]*Department of Physiology & Neuroscience, St. Georges University, Grenada, West Indies*

**Abstract**

Patients with idiopathic generalised epilepsy (IGE) typically have normal conventional magnetic resonance imaging (MRI), hence diagnosis based on MRI is challenging. Anatomical abnormalities underlying brain dysfunctions in IGE are unclear and their relation to the pathomechanisms of epileptogenesis is poorly understood.

In this study, we applied *connectometry*, an advanced quantitative neuroimaging technique for investigating localised changes in white-matter tissues *in vivo*. Analysing white matter structures of 32 subjects we incorporated our *in vivo* findings in a computational model of seizure dynamics to suggest a plausible mechanism of epileptogenesis.

Patients with IGE have significant bilateral alterations in major white-matter fascicles. In the cingulum, fornix, and superior longitudinal fasciculus, tract integrity is compromised, whereas in specific parts of tracts between thalamus and the precentral gyrus, tract integrity is enhanced in patients. Combining these alterations in a logistic regression model, we computed the decision boundary that discriminated patients and controls. The computational model, informed with the findings on the tract abnormalities, specifically highlighted the importance of enhanced cortico-reticular connections along with impaired cortico-cortical connections in inducing pathological seizure-like dynamics.

We emphasise taking directionality of brain connectivity into consideration towards understanding the pathological mechanisms; this is possible by combining neuroimaging and computational modelling. Our imaging evidence of structural alterations suggest the loss of cortico-cortical and enhancement of cortico-thalamic fibre integrity in IGE. We further suggest that impaired connectivity *from* cortical regions *to* the thalamic reticular nucleus offers a therapeutic target for selectively modifying the brain circuit for reversing the mechanisms leading to epileptogenesis.

## 1. Introduction

Idiopathic generalised epilepsies (IGEs) constitute nearly a third of all epilepsies and can manifest with typical absences, myoclonic jerks, and generalised tonic-clonic (GTC) seizures, alone or in varying combinations (Panayiotopoulos 2005; Berg et al. 2009). Several clinical IGE syndromes are traditionally distinguished (e.g., juvenile myoclonic epilepsy (JME); juvenile absence epilepsy; epilepsy with generalised tonic-clonic seizures on awakening (GTC-A)), however observational data suggests considerable overlap between these syndromes (Reutens and Berkovic 1995). The pathogenesis of IGE is not understood. The "idiopathic" modifier implies an unknown cause in contrast to, for example, epilepsies where an underlying structural cause is evident. Routine MR imaging is unremarkable.
There is strong empirical evidence for heritability (Berg et al. 2009), reflected in the recently proposed change in terminology to genetic generalised epilepsies (Berg et al. 2010). However, the IGE term is still widespread in non-clinical literature and will be used here.

---

[1]n.sinha2@newcastle.ac.uk
[2]peter.taylor@newcastle.ac.uk

Electrographically, seizures in IGE are characterised by simultaneous bilateral epileptiform discharges. However, on standard MR imaging, patients with IGE appear normal. While more advanced methods have identified some changes, these findings are inconsistent (Duncan 2005). It has long been suggested that there are shared pathophysiological mechanisms across the spectrum of IGE syndromes (Andermann and Berkovic 2001; Benbadis 2005). Therefore, investigation of mixed-syndrome IGE populations is important in identifying shared pathophysiology (Pitknen et al. 2016; Whelan et al. 2018). Moreover, examining brain structures beyond those prominent in the standard MR imaging is called for.

Although there is a paucity of studies on white-matter changes in IGE, some studies have demonstrated regional white matter alterations. In patients with JME, anterior thalamo-cortical radiations to frontal lobe networks have been consistently found abnormal (Deppe et al. 2008; O'Muircheartaigh et al. 2012; Keller et al. 2011; Liu et al. 2011). Additionally, motor networks, along with callosal abnormalities have also been found to be implicated (O'Muircheartaigh et al. 2011; Vollmar et al. 2012; Gong et al. 2017). For patients with GTCs, while (Liu et al. 2011) found no aberration in the white-matter, (Li et al. 2010) applied multiple analysis techniques and found abnormalities in parts of cerebellum but not elsewhere. However, (Zhang et al. 2011) and (Liao et al. 2013) detected abnormalities in the non-cerebellar regions by applying graph theory and found altered node topological characteristics in default mode network, mesial frontal cortex and in subcortical structures. Combining functional and structural connectivity, (Zhang et al. 2011) demonstrated loss in structure-function coupling and epilepsy duration. (Ji et al. 2014) found subtle reduction in lengths of commissural tract bundles connecting the anterior cingulate cortex and the cuneus bilaterally, which were also found to have altered functional connectivity. In childhood absence epilepsy (CAE), white-matter has been shown to be impaired by analysing specific structures (Luo et al. 2011; Yang et al. 2012) and also by applying network-based graph theoretical measures (Xue et al. 2014). Furthermore, (Qiu et al. 2016) found impairments in default mode network regions. In mixed syndrome IGE, aberrant increases in thalamic, precentral, and parietal areas have been shown (Groppa et al. 2012) in addition to differences in callosal and cortico-spinal tracts amongst others (Focke et al. 2013). On the other hand, (McGill et al. 2014) found no structural difference in patients with IGE. In summary, the existing white matter findings vary widely between studies, possibly due to different methods applied for detecting complex patterns of microstructural alterations.

Diffusion MRI Connectometry is a recently developed analytical method that is more sensitive to local structural differences in fibre pathways (Yeh et al. 2013a, 2016). This is fundamentally different from conventional whole-tract based approaches where end-to-end tracking between different regions is performed first and then differences in connectivity sought. A summary measure derived for the entire track may obscure localised pathological changes (Yeh et al. 2016; Keller et al. 2016). In contrast, connectometry first maps the difference in a parameter of interest at each fibre orientation in a voxel. It then connects the differences using fibre tracking, thus delineating an entire altered pathway (Abhinav et al. 2014b). The exact location and extent of the abnormality can be detected with high specificity. Connectometry has been applied in a number of studies investigating structural alterations such as in chronic stroke, amyotrophic lateral sclerosis, depression amongst others (Yeh et al. 2013a; Abhinav et al. 2014a). However, to our knowledge, connectometry has not been applied to examine white-matter changes in IGE.

Beyond the analysis of static properties of brain connectivity, computational models additionally allow the prediction of time-varying activity (Lytton 2008; Baier et al. 2012). Models simulate time series of cerebral activity, constrained by model parameters. In recent years, models constrained with parameters derived from clinical data have provided mechanistic insights into the pathophysiology of seizure genesis, maintenance, spread, and termination (e.g., Kramer et al. 2012; Jirsa et al. 2017; Sinha et al. 2017; Proix et al. 2017; Bauer et al. 2017). In the context of spike-wave seizures, models of thalamocortical interactions have been proposed (Destexhe 1998; Robinson et al. 2002; Breakspear et al. 2006; Marten et al. 2009; Taylor et al. 2014b). However, it is only recently that models of IGE have been constrained by subject-specific data (Taylor et al. 2013; Nevado-Holgado et al. 2012; Yan and Li 2013; Benjamin et al. 2012; Schmidt et al. 2014, 2016). Combining a dynamical model with subject-specific neuroimaging data permits exploration of possible mechanistic consequences of observed connectivity change (Taylor et al. 2014a; Proix et al. 2017).

The focus of this study is to elucidate white matter changes that are common to mixed-syndrome IGE, using diffusion MRI Connectometry, a potentially more sensitive method. We then explored possible mechanistic consequences of the connectivity changes in a computational dynamic model, constrained by our neuroimaging observations.

## 2. Methods

### 2.1. Participants

We studied a total of 32 subjects recruited by the New York University Comprehensive Epilepsy Centre. Our subject cohort included 14 patients with IGE (7 males and 7 females, age range 20.6-49.6 years, mean age 34 years). Patients were age and gender matched with 18 healthy control subjects (9 males and 9 females, age range 20.9-46.5 years, mean age 30.2 years). Details of functional connectivity change in this cohort have been previously reported (McGill et al. 2012, 2014). The details of patients are provided in Table 1 and information on control subjects is in Supplementary Table 1. Patients met the criteria for IGE and had no history of developmental delay or structural brain abnormalities. Standard diagnostic anatomical imaging studies were normal. Electrophysiological evaluation with interictal, and in most cases, ictal EEG demonstrated typical generalised epileptiform spikes. Patients with focal epileptiform discharges or focal slowing on EEG were excluded (McGill et al. 2012, 2014). Patients with IGE were classified according to the International League Against Epilepsy (ILAE) classification as having absence seizures (35.7%), myoclonic seizures (50%), generalised tonic-clonic seizures (85%), or combinations thereof as shown in Table 1. All people diagnosed with IGE were receiving active medical treatment at the time of study. All subjects gave their written informed consent to participate in this study, which was approved by the Institutional Review Board of NYU Langone School of Medicine.

Table 1: Patient information and seizure types

| | Demographics | | | Seizure types | | |
|---|---|---|---|---|---|---|
| Patient | Gender | Age | Epilepsy duration | Onset Age | Absence | Myoclonic | GTC |
| 1 | M | 49.6 | 30.6 | 19 | | × | |
| 2 | F | 48.7 | 31.7 | 17 | × | × | × |
| 3 | M | 47.4 | 31.4 | 16 | | | × |
| 4 | M | 27.5 | 7.7 | 19.8 | | × | |
| 5 | F | 32.1 | 12.1 | 20 | | | × |
| 6 | F | 21.7 | 3.6 | 18.1 | × | | × |
| 7 | F | 38.8 | 26.8 | 12 | | × | × |
| 8 | M | 27.7 | 26.2 | 1.5 | | | × |
| 9 | M | 35.6 | 26.6 | 9 | × | | × |
| 10 | M | 36.3 | 5.3 | 31 | | | × |
| 11 | F | 24.6 | 9.6 | 15 | × | × | × |
| 12 | M | 20.6 | 5.6 | 15 | | | × |
| 13 | F | 37.9 | 23.9 | 14 | | × | × |
| 14 | F | 27.6 | 17.8 | 9.8 | × | × | × |

### 2.2. Data acquisition and processing

All 32 subjects underwent scanning on a Siemens Allegra 3.0T scanner at New York University Centre for Brain Imaging. Participants had a T1-weighted MRI sequence optimised for grey-white matter contrast (TR = 2530 ms, TE = 3.25 ms, T1 = 1100 ms, flip angle = 7°, field of view (FOV) = 256 mm, matrix = 256 × 256 × 192, voxel size = 1 × 1.33 × 1.33 mm). Images were corrected for non-linear warping caused by non-uniform fields created by the gradient coils. All participants also had diffusion MRI scans. Diffusion-weighted echo-planar MRI were acquired by applying diffusion gradients along 64 directions ($b$-value = 1000 s/mm$^2$) with the following parameters during the 6 min 3s scan (TR = 5500 ms, TE = 86 ms, FOV = 240 mm, slice thickness = 2.5 mm, voxel size = 2.5 × 2.5 × 2.5 mm). Diffusion data were corrected for eddy current and motion artefacts using the FSL eddy correct tool (Andersson and Sotiropoulos 2016). We then rotated the $b$ vectors using the 'fdt-rotate-bvecs' tool (Jenkinson et al. 2012; Leemans and Jones 2009).

*2.3. Diffusion weighted imaging analysis*

*2.3.1. Tractography*

We analysed the data obtained from eddy corrected diffusion-weighted MRI and T1-weighted MRI in the DSI Studio (`http://dsi-studio.labsolver.org`) software pipeline. The diffusion data were reconstructed in standard space using q-space diffeomorphic reconstruction (QSDR) (Yeh and Tseng 2011) to obtain the spin distribution function (SDF) (Yeh et al. 2010). The diffusion sampling length ratio was 1.25 and the QSDR reconstruction yields maps of SDFs at 2 mm isotropic resolution. We applied an 8-fold orientation distribution function (ODF) tessellation with 5 peaks of the ODF, allowing for crossing fibers within voxels. For fibre tracking, we seeded the regions and the tracts were terminated when the quantitative anisotropy of the voxel through which the streamline entered was below 0.6*(Otsu's threshold). Otsu's threshold is calculated to give the optimal separation threshold that maximises the variance between background and foreground (Otsu 1979). We choose deterministic tractography due to the likelihood of it generating fewer false positive connections than probabilistic approaches. Streamlines with implausible lengths (> 300 mm and < 10 mm) and extreme turning angles (> 60 degrees) were excluded. Other parameters were as follows: step size: 1 mm, smoothing: 0, seed orientation: primary, seed position: subvoxel, randomise seeding: 0, direction interpolation: trilinear, tracking algorithm: runge-kutta order 4, streamlines threshold: 1,000,000. We assessed the integrity of tracts by analysing the generalised fractional anisotropy of the tract profiles. The gFA measure is a generalised version of the widely used FA measure of white matter integrity (Tuch 2004).

*2.3.2. Connectometry*

We investigated the structural alterations in white matter pathways using diffusion MRI connectometry to compare patient and control groups (Yeh et al. 2016). Connectometry permits analysis of the white-matter characteristics locally in contrast to the conventional end-to-end, whole-tract approaches. It has been shown that when the structural change involves only a segment of the white-matter pathways, global connectivity and tractography based approaches have a much lower sensitivity in capturing the regional variability in the white matter as compared to connectometry (Yeh et al. 2013a; Faraji et al. 2015; Yeh et al. 2016). The overall procedure of the connectometry analysis is illustrated in Figure 1.

**[FIGURE 1]**

First, we created a connectometry database of all 32 subjects (14 patients and 18 controls) with generalised fractional anisotropy (gFA) as the index of interest (Figure 1(a,b)). In creating the connectometry database we used the default Human Connectome Project HCP842 atlas as the common atlas from which the local fibre directions are sampled to create the local connectome matrix. To aid reproducibility, we provide an anonymised version of the Connectometry database we used in this study on (*link to be generated upon acceptance*).

Second, we set-up the group connectometry analysis in DSI studio to correlate the local connectome matrix with age, gender, and subject category (patient or control) in a multiple linear regression model while selecting the subject category as the study variable. In this step, we also assigned a seed region to study the regional differences of the local connectomes emanating from that region (Figure 1(c)). Specifically, we seeded (i) the whole brain to visualise the consistent trend of connectome alterations,

(ii) all regions defined in the JHU white-matter atlas excluding the regions in cerebellum and brainstem, and (iii) corpus callosum and thalamic regions from the freesurfer Destrieux atlas. All the regions studied along with their anatomical locations are shown in Supplementary Figure 1.

Third, we ran connectometry analysis for tract length threshold varying from 20*mm* to 60*mm* in increments of 2*mm* (voxel resolution) and *t*-score threshold varying from 1.5 to 2.5 in increments of 0.05. At each point on this two dimensional grid, local connectomes for every seed region under study are delineated with positive and negative associations of gFA and the subject category (Figure 1(d,e)). The local connectomes were tracked using a deterministic fibre tracking algorithm (Yeh et al. 2013b), track trimming was iterated once, and the seed count was set to 10000. These are the default settings for fibre tracking in connectometry analysis from the 9 August 2018 version of DSI Studio. Connectometry analysis controls for the true positive and false positive rate by comparing the positive and negative associations of gFA with a null distribution by randomly permuting the group labels. Further details of this procedure can be referred in the section on "Local connectomes and their statistical inference" in Yeh et al. 2016 and also in the method section of Yeh et al. 2013a. In our study, we set the permutation count to 1500 which is reasonably high and a pragmatic choice for speed.

Fourth, we quantified the output of the connectometry analysis at each point of the two-dimensional grid for every seeded region of interest. We computed the mean gFA of the voxels traversed by the altered tracts detected from the connectometry analysis for each subject. These mean gFA values were exported in MATLAB and effects of age and gender were regressed out to compute the residuals. The residual gFA were compared between the patient and the control groups by applying a non-parametric Wilcoxon rank sum test (Figure 1(f)). We corrected the *p*-values for multiple comparison by applying Benjamini-Hochberg false discovery rate procedure at 5% significance level (Figure 1(h)) and then applied a binary threshold at $p = 0.05$ (Figure 1(i)). Identifying bundles of adjacent, altered, streamlines is also important in avoiding Type-I errors. Therefore, we noted the number of streamlines detected at each point of the two-dimensional grid (Figure 1(g)). An arbitrary binary threshold of $n = 100$ was applied to visualise the range of tract length and *t*-score values for which the connectometry analysis identified at least 100 streamlines as aberrant (Figure 1(j)). This binary matrix was multiplied element-wise with the binary matrix generated from the FDR corrected *p*-values (Figure 1(k)). Thus, we determined a range of tract lengths and *t*-scores where there exist significant microstructural alterations in at least 100 tract bundles for each seeded region.

*2.4. Dynamical model*

To investigate how aberrations in connectivity may lead to seizure dynamics, we related our data-driven neuroimaging findings to an established mathematical model which simulates dynamic thalamocortical interactions. The model is illustrated schematically in Figure 1(l). This neural population model is based on the Wilson-Cowan formalism (Wilson and Cowan 1972) which is one of the best-studied population level models (see e.g., Borisyuk et al. 1995; Destexhe and Sejnowski 2009; Wang et al. 2012; Meijer et al. 2015).

In this neural population model, the cortex is treated as one subsystem and the thalamus as the other. The cortical subsystem is composed of excitatory pyramidal (PY) and inhibitory interneuron (IN) populations. The thalamic subsystem includes variables representing populations of thalamo-cortical relay cells (TC) and neurons located in the reticular nucleus (RE). The connection schemes, shown by the arrows in Figure 4(b) between various cell populations, have the following interpretation: the PY variable is self-excitatory and also excites the IN population. In addition, PY excites the TC and RE cells of thalamus. IN inhibits local cortical PY cells only. Direct thalamic output to the cortex comes only from the excitatory TC populations to the PY populations. Intra-thalamic connectivity is incorporated as the TC cells having excitatory projections to RE, which in turn inhibit the TC population along with self-inhibition of RE. The dynamics of the model are governed by the following differential equations inspired by the model in Taylor et al. 2014b:

$$\tau_1 \frac{dPY}{dt} = -PY + S(C_1 PY - C_3 IN + C_9 TC + h_{py}) \quad (1)$$

$$\tau_2 \frac{dIN}{dt} = -IN + S(C_2 PY + h_{in}) \quad (2)$$

$$\tau_3 \frac{dTC}{dt} = -TC + C_7 PY - C_6 RE + h_{tc} + \alpha N(t) \quad (3)$$

$$\tau_4 \frac{dRE}{dt} = -RE + C_8 PY - C_4 RE + C_5 TC + h_{re} \quad (4)$$

where $h_{py,in,tc,re}$ are external input parameters which can be interpreted as the general excitability level, as they determine how much input is required to activate a population (Wilson and Cowan 1972; Amari 1977). $\tau_{1,\cdots,4}$ are time scale parameters, $C_{1,2,\cdots,9}$ are connectivity parameters between the cortical and thalamic subsystems, $N(t)$ is normally distributed noise input with zero mean and unit standard deviation; $\alpha$ modulates the noise amplitude. The noise term was added to the TC population following previous modelling literature of the thalamo-cortical loop, and represents non-specific ascending noise input from the brain stem (Robinson et al. 2002; Breakspear et al. 2006; Marten et al. 2009; Taylor et al. 2014b). The formalism for the noise term used here is adopted in Wang et al. 2014, 2017. $S[\cdot]$ is the sigmoid activation function defined as follows:

$$S[x] = \frac{1}{(1+\exp(-a(x-\theta)))} \quad (5)$$

where, $a = 1$ is the steepness of the sigmoid function, and $\theta = 4$ is the $x$ offset Wang et al. 2012, 2014.

We simulated the model numerically in MATLAB (The MathWorks, Natick, MA). We computed the solutions of the deterministic model (i.e., noise term $N(t)$ in equation 3 is set to zero) by using the ode45 solver. For the stochastic model, in which the noise term is included to simulate seizure transitions, we implemented the equations as a stochastic differential equation, and used the Euler-Maruyama solver with step size $\delta = (1/15000)s$. Tutorials and MATLAB codes for numerically simulating SDEs are explained in Higham 2001. The simulated EEG dynamics in the model correspond to the activity of PY after filtering out the DC component with a high pass filter at 1 Hz (see, Jirsa et al. 2014; Wang et al. 2017). We reconstructed a regime where the spike-wave attractor was bistable to the fixed point by understanding the bifurcation structure of our model using a phase space approach through reconstruction of subsystems which, when operating on different timescales can be combined (Fenichel 1979; Wang et al. 2012; Jirsa et al. 2014). The parameter values incorporated in the model are provided in Supplementary Table 2. Their choice was guided by previous works that highlighted the necessary bifurcation structures (Wang et al. 2012; Taylor et al. 2014b).

*2.5. Discriminatory decision boundary*

We used a logistic regression model to infer a linear decision boundary that optimally discriminates IGE patients and controls. Recognising the two different patterns of alterations, i.e., *decreases* in cortico-cortical tracts and *increases* in thalamo-cortical tracts (see, Results), we chose to derive two features from our data. Specifically, we considered the mean gFA of voxels traversed by the abnormal thalamo-cortical tracts as one feature. Similarly, we incorporated the mean gFA of all the voxels underlying the abnormal cortico-cortical tracts as the second feature. Age and gender were regressed out from both the features using a multiple linear regression model. Therefore, our feature space can be represented as $A \in R^{n \times m}$, where $m = 2$ indicates the total number of features and $n = 32$ denotes the total number of subjects in our study.

The linear decision boundary between IGE patients and controls was determined using a logistic regression model by minimising the following regularised objective function:

$$\min_{x} \sum_{i=1}^{n} \log(1 + e^{(-y_i(x^T a_i + c))}) + \frac{\rho}{2} \|x\|_2^2 \quad (6)$$

where, $y = (y_1, y_2, \cdots, y_n)$ is the $n$ dimensional vector representing the two groups (0 for healthy subjects and 1 for patients with IGE); $a_i^T$ denotes the $i$-th row of feature space $A \in R^{n \times m}$; $c$ is the scalar intercept; and $\frac{\rho}{2}\|x\|_2^2$ is the $l_1$ (ridge) regularisation term. We optimized the cost function 6 by applying stochastic gradient descent algorithm as implemented in the *fitclinear* method in MATLAB in which $\rho$ is set to $1/n$. Optimising the cost function in equation 6 with the entire data, we computed the discriminatory decision boundary between patients and controls.

*2.6. Statistical analysis and visualisation*

We applied the non-parametric Wilcoxon rank sum test for comparing the white matter differences between the patients and controls. We corrected for multiple comparisons by applying Benjamini-Hochberg false discovery rate correction at a significance level of 5% (Benjamini and Hochberg 1995). Results were declared significant at $p < 0.05$. To illustrate the anatomical location of the observed differences in the white matter tracts, we reconstructed the section-wise *t*-score plots. Cohen's d measures the standardised difference between two means (Cohen 1988). Therefore, we computed the Cohen's *d*-score to measure the effect size of alteration of white matter differences between the patient and control groups. We created the scatter plots using the UniVarScatter function in MATLAB (https://github.com/GRousselet/matlab_visualisation).

## 3. Results

Our results are presented in three main sections. First, we demonstrate *microstructural* connectivity alterations in the white matter tracts of the default mode network (DMN), complementing previous demonstrations of DMN resting-state fMRI *functional* connectivity alterations in the same subjects (McGill et al. 2012, 2014). Second, we illustrate the white matter aberrations found in patients after analysing the entire white-matter structures of the two groups. Finally, we demonstrate how alterations in connectivity affect IGE seizure dynamics in the computational model.

*3.1. Microstructural changes in the default mode network*

Recent studies have highlighted that anatomical connectivity between brain regions underpins functional connectivity (Chu et al. 2015; Honey et al. 2009). Impaired functional connectivity in the default mode network has been shown previously by McGill et al. 2012 for the same idiopathic generalised epilepsy patients as in our study. Specifically, McGill et al. 2012 demonstrated that with the seed placed in the posterior cingulate cortex (PCC), its functional connectivity to the ventral part of the medial prefrontal cortex (MPFC) is reduced in IGE patients (results reproduced in Figure 2(a)). These regions are located within the PCC and MPFC, two of the prominent nodes of the default mode network. Previous studies have identified the cingulum tracts as the only direct anatomical connection between these two regions of the default mode network (see, Table 1 in Van Den Heuvel et al. 2009). Here, we investigate if the demonstrated alterations in functional connectivity have a basis in the corresponding altered structural connectivity.

[FIGURE 2]

First, we delineated all the cingulum tracts connecting PCC and MPFC by performing conventional end-to-end tractography. These tracts are shown in Figure 2(b). Next, for detecting any partial abnormalities in these tracts, we performed connectometry by setting the cingulum region from the JHU white-matter atlas as seed. The tract profile of cingulum fibres are colour coded in accordance with their *t*-scores that quantifies how different the tracts are with respect to controls. Across a wide range of *t*-score and tract length thresholds, we found that in IGE patients there is a significant loss of tract integrity in segments of the cingulum tracts. In Figure 2(b), we have shown these alterations at a representative *t*-threshold of $t = 1.94$ and length threshold of $l = 30mm$. Significant reduction at these thresholds are quantified in the corresponding box plot ($p = 0.01$, $d = 0.90$) in Figure 2(b). Consistent replication of this trend across a wide range of tract length and *t*-score thresholds is illustrated in Supplementary Figure 1(a). These results give an indication that the loss of white-matter integrity in the cingulum tracts demonstrated here might be a plausible basis for the reduced functional connectivity between the regions connected by it (i.e., PCC and ventral MPFC in McGill et al. 2012).

Note that the application of conventional whole-tract based approach in which summary statistics is derived by averaging gFA across the entire tract profile (i.e., ignoring partial abnormalities), failed to detect any significant difference as illustrated in Supplementary Figure 5. As expected, this is mainly because the partial abnormalities of the cingulum tracts are eclipsed by the normal segments (shown in white in Figure 2(b)).

*3.2. White matter structures with altered connectivity*

We identified four pairs of structural connections that were consistently altered between patients and controls across a large range of *t*-score and tract length thresholds (Figure 3(c)) . The cingulum tract aberrations discussed in the previous section are shown again in Figure 3 alongside other white-matter abnormalities detected. Apart from cingulum tracts, we also found that the integrity of the white matter tracts in the column and body of fornix was significantly compromised in patients. Tracts from fornix detected at an illustrative threshold of $t = 1.94$ and length $= 30mm$ from connectometry analysis are shown in Figure 3(a). As depicted by the box plot in Figure 3(b), gFA residuals averaged across the tract and compared between the two groups is significantly reduced ($p = 0.02$, $d = 0.87$) in IGE patients.

[FIGURE 3]

Similarly, we detected substantial microstructural abnormality in the white matter segments of the bilateral superior longitudinal fasciculus. We found that the fibre integrity of superior longitudinal fasciculus was compromised in patients with IGE for a wide range of thresholds. In Figure 3(a), we plot the abnormal segments of superior longitudinal fasciculus at exemplary threshold values of $t = 1.94$ and length $= 30mm$. Significant reduction of mean gFA across the tract in these segments for IGE patients at $p = 0.002$ and $d = 1.34$ are shown in the box plot in Figure 3(b).

In contrast, we found that the white matter tracts forming part of the bilateral cortico-thalamic radiations between the thalamus and the precentral gyrus, have an enhanced structural integrity in patients with IGE. This was evident from the connectometry analysis where these tracts showed an increased gFA association in IGE patients across a wide range of t-score and tract length thresholds. In Figure 3(a), these tracts are plotted at an illustrative threshold value of $t = 1.94$ and length $= 30mm$. Aberrant increase of gFA associated with these thalamo-cortical radiations are quantified in the box plot (Figure 3(b)) at $p = 0.005$ and $d = -1.0$ by comparing the mean gFA underlying these tracts between IGE patients and controls.

In summary, we have demonstrated that the tracts in bilateral cingulum, fornix, and superior longitudinal fasciculus have reduced gFA, but parts of bilateral thalamo-cortical radiations have increased gFA in patients. This pattern of reduced structural integrity in cortico-cortical connections and enhanced structural integrity in thalamo-cortical connections was confirmed upon analysing each white matter region of the JHU atlas and a few other ROIs (for completeness) from Destrieux atlas which are illustrated in Supplementary Figure 1. Visually also this pattern of alteration is evident from seeding the whole-brain in connectomery analysis illustrated in Supplementary Figure 2. Taken together, our findings suggest that in idiopathic generalised epilepsy, there is a trend towards a loss of white-matter integrity in cortico-cortical connections, whereas thalamo-cortical connections tend to be abnormally enhanced. In Figure 4(a), we show the optimal decision boundary for discriminating patients and controls. Note that the patients (controls) with increased (decreased) fibre integrity in cortico-thalamic tracts and decreased (increased) fibre integrity in cortico-cortical tracts are above (below) the positively-sloped decision boundary. We incorporate this information to inform the analysis of the dynamical model in section 3.3.

*3.3. Predicting epileptogenesis mechanisms due to connectivity alterations*

Diffusion MRI can tell us which tracts are aberrant in patients, however, it does not tell us their direction, or how this may lead to epileptogenesis mechanistically. We therefore incorporate the structural white-matter connectivity changes demonstrated in the previous section into a dynamical model to examine whether such changes could contribute to epileptogenesis.

*3.3.1. Model dynamics*

We implemented a neural population model with cortical and thalamic subsystems illustrated in Figure 4(b) to investigate the mechanisms of seizures due to connectivity alterations. We are specifically interested in the four parameters of this model which reflect our neuroimaging findings: (i) cortico-cortical (PY→PY) connectivity C1, (ii) cortico-thalamic relay nuclei (PY→TC) connectivity C7, (iii) cortico-reticular (PY→RE) connectivity C8, and (iv) thalamo-cortical (TC→PY) connectivity C9. These parameters are highlighted in Figure 4(b).

**[FIGURE 4]**

In Figure 4(c), we have charted the model dynamics for variations in cortico-cortical parameter (C1) with respect to cortico-thalamic parameters (C7, C8, and C9). In the blue regime of the parameter space, the system resides in the background state of fixed point dynamics. This represents seizure free activity, similar to most models (e.g., Breakspear et al. 2006; Wang et al. 2012; Jirsa et al. 2014; Sinha et al. 2017). In the orange regime, however, the model exhibits spike-wave dynamics. Spike-wave discharges are typically noted as the clinical marker of pathological seizure activity in many types of IGE syndrome. Therefore, the orange regime of the parameter space represents the pathological region where seizures would ensue. In addition, the model is also capable of producing monostable fast oscillatory dynamics in the region of 20-30 Hz (grey regions), which could represent additional seizure patterns (Wang et al. 2012, 2017). For the purpose of this study, we consider both the orange and grey region to be epileptogenic. Note that we do not assume a specific dynamic mechanism by which seizures occur (i.e., if the seizure occurs due to some slow parameter change; or due to spontaneous noise-driven transitions, which are possible in bistable states (Da Silva et al. 2003; Baier et al. 2012; Wang et al. 2014)). Rather we only wish to highlight parameter regions that can support seizure dynamics in principle (in other words, where a seizure attractor exists or co-exits). Our model also features a bistable region, where fixed-point and SWD attractors coexist (Supplementary Figure 4), thus we are not excluding any particular mechanism of ictogenesis. In our modelling framework, connectivity parameters for patients would assume a value such that the dynamics are placed in the epileptogenic (orange, grey) region where seizures occur whereas, the connectivity parameter for controls will be in seizure free (blue) region. Nonetheless, we have shown the subdivisions of the parameter space with exact dynamics in the bifurcation plot, capturing the minima and maxima of simulated time-series for representative parameter values in Supplementary Figure 4.

An example of clinical seizure recorded from the EEG of an IGE patient is shown in Figure 4(e). Note the spike-wave discharges (SWD) occurring during seizures. In addition to the clinical spike-wave dynamics, we also show SWD-like signals generated by our model. As evident, in the model, many key features of clinical spike-wave seizures can be reproduced, including repeating SWD oscillations, morphology of SWD, change in frequency, and fast spike followed by a slow wave.

*3.3.2. Connectivity alteration inducing seizure dynamics*

The tracts obtained from diffusion MRI analysis are not directed, even though fibre tracts in the human brain are directed. To investigate this, in the model we incorporated directional thalamo-cortical connectivity parameters, which may allow us to predict specific thalamic connections that could lead to seizure dynamics. We scanned each directed thalamic connectivity parameter in the model (C7, C8, C9) with respect to the cortico-cortical connectivity parameter (C1) and determined the bifurcation orientation separating the normal and pathological dynamics as shown in Fig. 4(c). With connectivity alterations unconstrained in the model, a number of parameter changes in e.g., C1, C7, C8, C9 can lead to epileptogenic dynamics. Therefore, we constrain the model by validating the orientation of bifurcation in parameter space with the orientation of decision boundary between patients and controls which we previously detected from our imaging data (Fig. 4(a)). Using imaging data is crucial to constrain the parameter space of the model within biologically informed boundaries. These constraints facilitate making model-based predictions while also maintaining concordance with the imaging observations.

Imaging analysis in previous sections revealed that the patients have a decreased cortico-cortical connectivity and increased thalamo-cortical connectivity. Thus, the decision boundary has a positive slope with patients (controls) occupying the space above (below) the decision boundary. This observation concords strikingly with the bifurcation orientation between cortico-thalamic connections to reticular nucleus (PY→RE, C8) with respect to cortico-cortical connections (PY→PY, C1) as shown in Figure 4 (panel c(i)). In this panel, the seizure and seizure free regime are separated with a positive slope with the former above the later. This, however, is not the case in Figure 4 (panel c(ii)) for C1 vs. C9 and in Figure 4 (panel c(iii)) for C1 vs. C7. In both these panels, the bifurcation orientation between the region supporting spike-wave and fixed-point dynamics have a negative slope that does not agree with the dMRI data. Therefore, the model implicates the role of thalamic reticular nuclei in seizures which was not otherwise observable from neuroimaging analysis alone. Note the bifurcation orientation between cortico-reticular connections (C8) with respect to cortico-cortical connections (C1) is preserved with slight perturbations in parameter C7 and C9 as shown in Figure 4(d), suggesting robustness of the result presented.

An exemplary point representing the control population in the region of the normal seizure free dynamics is shown by a circle in green. The model predicts various routes that may place the healthy dynamics in controls to the pathological regime where spike-wave dynamics manifest. As illustrated by the arrows, reduction of cortico-cortical (C1) parameter (purple line), increase in C8 parameter (orange line), or a combination of the two (cyan line) can induce a bifurcation to spike-wave dynamics in the model.

Putting together our model predictions, it appears that the localised alterations in white matter structures contribute to the mechanism of seizures in IGE. After model validation (matching the decision boundary from the dMRI data to our bifurcation orientation), our proposed model suggests that epilepsy manifest due to enhancement of white matter connections *from* cortex *to* reticular nuclei coupled with reduction in cortico-cortical connections. Additionally, the model predicts that reducing the cortico-reticular connectivity can place the model in the normal seizure free regime again, thus offering a target which may be of therapeutic value in IGE.

## 4. Discussion

The main objectives of this study were a) to determine any anatomical abnormalities in white-matter structures in patients with mixed syndrome IGE, and b) to suggest plausible mechanisms of the pathogenesis in IGE.

We applied diffusion MRI Connectometry to identify localised structural connectivity abnormalities not evident using conventional whole-tract based approaches (Figure 2(a, b)). In doing so we identified a likely anatomical substrate for the reduced DMN functional connectivity already demonstrated in the same subjects by McGill et al. 2012.

Analysing the whole-brain white matter structure, we found that patients have decreased cortico-cortical connections consistently in parts of cingulum tracts, fornix, and superior longitudinal fasciculus. However, patients have increased connection strength between thalamic and precentral areas. To understand potential mechanisms, we incorporated our imaging findings into a computational model of thalamocortical interactions. Our modelling framework suggests increased cortico-reticular connectivity coupled with loss of cortico-cortical strength mechanises epileptogenesis in IGE. To our knowledge, this is the first study that combines diffusion imaging with computational modelling to underscore the pathophysiology of IGE being due to impaired white matter structure underlying cortical regions and cotico-thalamic projections of reticular nucleus.

### 4.1. Methodological considerations

In terms of cohort, our sample size of 14 patients is small; however, this is comparable to several previous studies of diffusion MRI analysis in IGE (Lee et al. 2014; Focke et al. 2013; Liu et al. 2011), and reflects the fact that diffusion weighted imaging is not routine in the clinical management of patients with IGE. Our patients were also not drug nave; therefore, it is unclear how our results may be confounded with the effects of medication. A longitudinal study of drug naive cohort is required to disentangle potential drug effects. Finally, we did not detect any significant correlation between clinical parameters (i.e., age of onset, epilepsy duration) and altered white-matter structure (which is perhaps again due to the small sample size of our data). Therefore, how the alterations of tracts change over time in patients remains unclear.

Different approaches can be applied to analyse diffusion MRI. Track based spatial statistics (TBSS) has been widely applied in various studies (Smith et al. 2006), including in epilepsy (Focke et al. 2013; Li et al. 2010; O'Muircheartaigh et al. 2011; Groppa et al. 2012; McGill et al. 2014; Lee et al. 2014; Gong et al. 2017). TBSS is a skeleton-based approach in which a mean FA skeleton is constructed and compared between the groups. While TBSS is an automated approach which has overcome some of the drawbacks of VBM based approaches (such as smoothing, alignment, and reproducibility issues), it still has some limitations. Summarising FA maps into a mean skeleton incurs loss of data; this may not make use of the information from the crossing fibres (Abhinav et al. 2014b). These problems have been overcome in diffusion MRI Connectometry which uses multiple fibre skeleton (multiple fibre orientations per voxel) to sample the diffusion quantities on ODFs. It incorporates $q$-space diffeomorphic reconstruction, a model-free reconstruction approach based on generalised $q$-sample imaging which has been shown to resolve the crossing fibre problem efficiently (Yeh et al. 2010, 2011, 2013b; Yeh and Tseng 2013; Abhinav et al. 2014b).

*4.2. Converging evidence of connectivity dysfunction*

The role of the thalamus in the pathogenesis of epileptic seizures has been long recognised (Gloor 1979; Niedermeyer et al. 1969; Avoli 2012). Simultaneous EEG-fMRI analysis has shown bilateral thalamic activation and deactivation of default mode network (Gotman et al. 2005; Moeller et al. 2010). This has further been corroborated with Positron Emission Tomography (PET) studies in which focal increases in thalamic blood flow have been reported (Prevett et al. 1995). Although a decrease in thalamo-cortical functional connectivity has been suggested between thalamus and frontal cortex (Kim et al. 2014), we did not detect any differences in the corresponding anatomical connectivity (Supplementary Figure 3), nor was it detected in McGill et al. 2014. Our findings of bilateral increased thalamo-cortical connectivity are in particular agreement with the study of Groppa et al. 2012 who also demonstrated alterations to FA in the thalamus and juxtacortical precentral areas.

The thalamic reticular nucleus is a part of circuitry normally responsible for generating healthy oscillations, such as sleep spindles (Fuentealba and Steriade 2005). Several studies have shown that pathological use of this circuitry results in manifestation of generalised spike-wave discharges (Gloor 1968; Meeren 2002; Meeren et al. 2005; van Luijtelaar and Sitnikova 2006; Huguenard and McCormick 2007; Beenhakker and Huguenard 2009; Lacey et al. 2012). Recently, EEG-fMRI analysis performed on IGE patients during early sleep stages revealed enhanced functional connectivity between thalamus and somatomotor region (which include precentral gyrus), amongst other regions (Bagshaw et al. 2017). These reports essentially complement our finding on abnormally high cortico-reticular anatomical connectivity as one of the contributory structures involved in mechanising epilepsy. We hypothesise based on our findings and from our model that abnormal input to the reticular nucleus from cortex can alter its normal inhibitory mechanism to generate pathological seizure activity.

We observed a reduction in white matter connections including the superior longitudinal fasciculus (SLF), fornix, and cingulum. SLF alterations were previously reported by Focke et al. 2013 who also found reductions in FA in a cohort of 25 patients. Similarly, a study by Lee et al. 2014 demonstrated microstructural alterations in thalamo-cortical and cortical-cortical connections in a cohort of 14 patients using diffusion kurtosis imaging. Furthermore, Liu et al. 2011 showed FA reductions in the SLF in a group of 15 patients with JME. The study of Liu et al. 2011 is in further agreement with ours since they also demonstrated FA reductions in regions of fornix and cingulum.

While one may expect a loss of structure with loss of function, which is also evident from our results on compromised cingulum fibres, the exact relationship on how anatomy constrains function and vice-versa is unclear. However, Honey et al. 2009 have shown that the relationship between structural and resting state functional connectivity is stronger in DMN than across the entire cerebral cortex. To some extent, interregional distance as well as direct or indirect connectivity have been suggested to be additional contributory factors (Chu et al. 2015; Honey et al. 2009). In this study, examining the cingulum tracts–identified previously as the only direct anatomical pathway between PCC and MPFC (Van Den Heuvel et al. 2009)–is hypothesis driven. We motivate future studies to include other contributory factors such as interregional distance and indirect connections, to further explore the structure-function relationship between PCC and MPFC in DMN.

By combining a dynamical model with imaging findings, we have shown evidence of how the white matter alterations in cortico-cortical and cortico-reticular structures may mechanise seizures. This is one of the key novelties of our study. With regard to the hotly debated topic of where the seizures manifest first, in cortex or in thalamus, we do not resolve that in this study as it would require simultaneous investigation of function and structure. However, there is a general consensus that the pathological mechanisms in generalised epilepsies are due to (i) increases in thalamo-cortical activity and (ii) broad decreases (with some possible regional increase) in cortico-cortical interactions (Blumenfeld 2002, 2003; see review by Duncan 2005).

*4.3. Model consideration*

We investigated the mechanism of epilepsy genesis in a two-subsystem neural population model of thalamo-cortical interactions. Note that epilepsy genesis (an increased propensity for seizures) is conceptually different to seizure genesis (mechanism underlying a specific transition to a seizure state), which we do not investigate here (Pitkänen and Engel 2014). The physiological basis of the model is based on the experimental evidence of connectivity shown by Pinault and O'brien 2005 and the references therein. Therefore, the resulting dynamics ensuing in coupled neural populations makes the computer model an apt framework for studying the effect of connectivity alteration between cortex and thalamus–as adopted in several other studies (Fan et al. 2016, 2017; Chen et al. 2017). Other models of seizure discharges comparable to the model incorporated here are: (Destexhe 1998) at microscopic scale of individual neurons, (Robinson et al. 2002) at a macroscopic scale of neural populations, along with phenomenological models (Jirsa et al. 2014), spatially extended models (Goodfellow et al. 2011; Taylor et al. 2012) amongst others (Bhattacharya et al. 2016; Breakspear et al. 2006; Marten et al. 2009; Yousif and Denham 2005, see review by Lytton 2008). The mechanism of SWD generation in these models may share some similarities, e.g., most models utilise the slow-fast bursting mechanism (Izhikevich 2000). Recently, a single framework of diverse bursting patterns has been suggested, enabling a unified classification of the mechanism and parameter changes leading to SWD in each model (Saggio et al. 2017). Using such a framework would be the next step to compare models and understand the full implication of our highlighted connectivity changes.

## 5. Concluding remarks

Case history and visual inspection of epileptiform activity on EEG are currently the standard tools for clinical diagnosis of IGE. Often case histories are unreliable and it is not unusual to get confronted with EEG negative cases (Smith 2005). Misdiagnosis incurs high cost (Juarez-Garcia et al. 2006), calling for the development of diagnostic biomarkers (Pitknen et al. 2016). Some efforts to devise alternative diagnostic tools based on resting-state EEG have been made (Schmidt et al. 2016). Furthering these attempts, we have identified non-invasive, MRI-based markers for discriminating patients with IGE. Structural changes are likely to occur at a much larger time-scale as compared to the dynamical properties of electrographic activity. Therefore, in a clinical setting, our discriminatory framework may be useful as an anatomical biomarker, although reproduction in a separate dataset would be important to verify its generalisability.

In conclusion, we found that patients with IGE have anatomical abnormalities in white-matter thalamo-cortical and cortico-cortical connections which are strikingly bilateral. We have demonstrated how a computational model can enable us to move beyond statistical observations in data to suggest a possible mechanism of IGE manifestation. Our analysis suggests the importance of increased directed connectivity *from* cortex *to* the thalamic reticular nuclei. The observed change creates a bistability in the network dynamics of our model – permitting the occurrence of occasional pathological epileptiform discharges. Taken together, our work may be of clinical interest for diagnostics, and the mechanistic insight suggests specific structural targets for the next generation of therapies in IGE.

# 6. Acknowledgement

NS was supported by Research Excellence Academy, Newcastle University, UK. PNT was supported by Wellcome Trust (105617/Z/14/Z and 210109/Z/18/Z). YW was supported by Wellcome Trust (208940/Z/17/Z). We thank Joe Necus, Gabrielle Schroeder, Frances Hutchings, Richard Rosch, Gerold Baier, and Rhys Thomas for discussions.

# 7. Supplementary Information

*Supplementary Figure 1.* Connectometry analysis upon seeding pre-defined regions of interest.

*Supplementary Figure 2.* Connectometry analysis upon seeding the whole brain.

*Supplementary Figure 3.* Thalamo-frontal white-matter integrity is preserved in IGE patients.

*Supplementary Figure 4.* Bifurcation diagram illustrating detailed model dynamics.

*Supplementary Figure 5.* Whole-tract approach obscures significant local differences in white matter.

*Supplementary Table 1.* Information on control subjects.

*Supplementary Table 2.* Values of the parameters incorporated in the model.

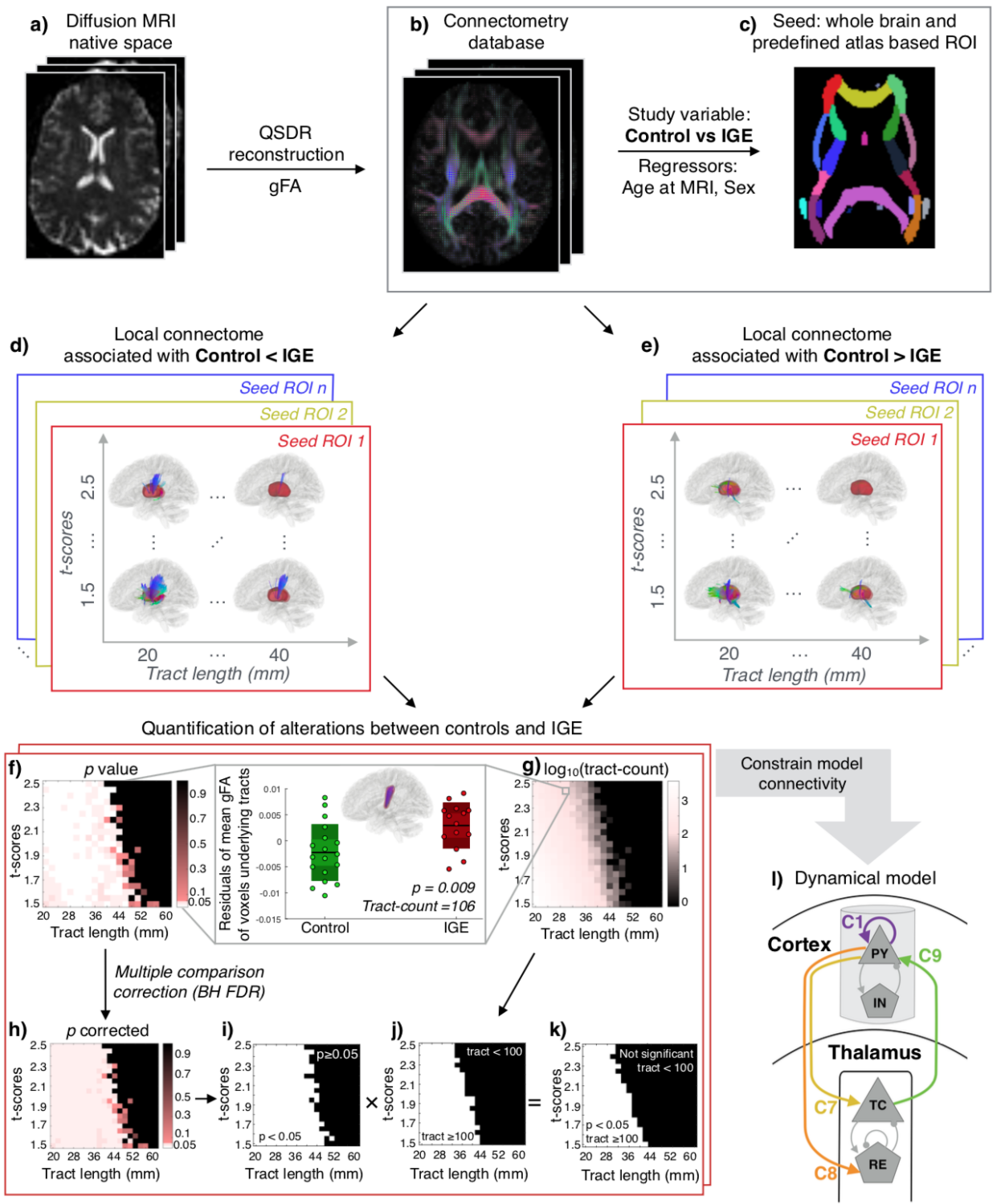

Figure 1: **Overall procedure.** (a-b) Diffusion MRI of each subject in the native space was reconstructed to the standard space for creating a connectometry database using QSDR reconstruction with gFA as the index of interest. (c) Group connectometry analysis was performed to study the local alterations in the white-matter tracts between controls and IGE patients using a multiple regression model with age at MRI acquisition and sex as covariates. Local connectome differences were studied for the whole brain seeding and seeding predefined atlas based regions. (d-e) For each seeded region of interest (ROI), connectometry was performed for a wide range of tract lengths and $t$-thresholds as represented by the two-dimensional space. At each point of the two-dimensional space, the local connectomes associated with the reduction and increase in IGE were delineated. (f-g) Output of connectometry were quantified by computing the significant mean gFA differences defined by the $p$-value between the groups and the tract count. (h) $p$-values were corrected for multiple comparisons using Benjamini-Hochberg FDR and a binary threshold was applied at 0.05 in (i). Similarly, on the tract count matrix, a binary threshold of 100 was applied to obtain a matrix shown in (j). The matrices in (i) and (j) were multiplied at each point to result in a two-dimensional space in (k) that highlights a region (in white) where the difference between the groups are significant with a minimum of 100 tracts detected. (l) Schematic of the thalamo-cortical model. Tract alterations from the imaging data were incorporated in the model to constrain the thalamo-cortical (C7, C8, C9) and cortico-cortical (C1) connectivity parameters.

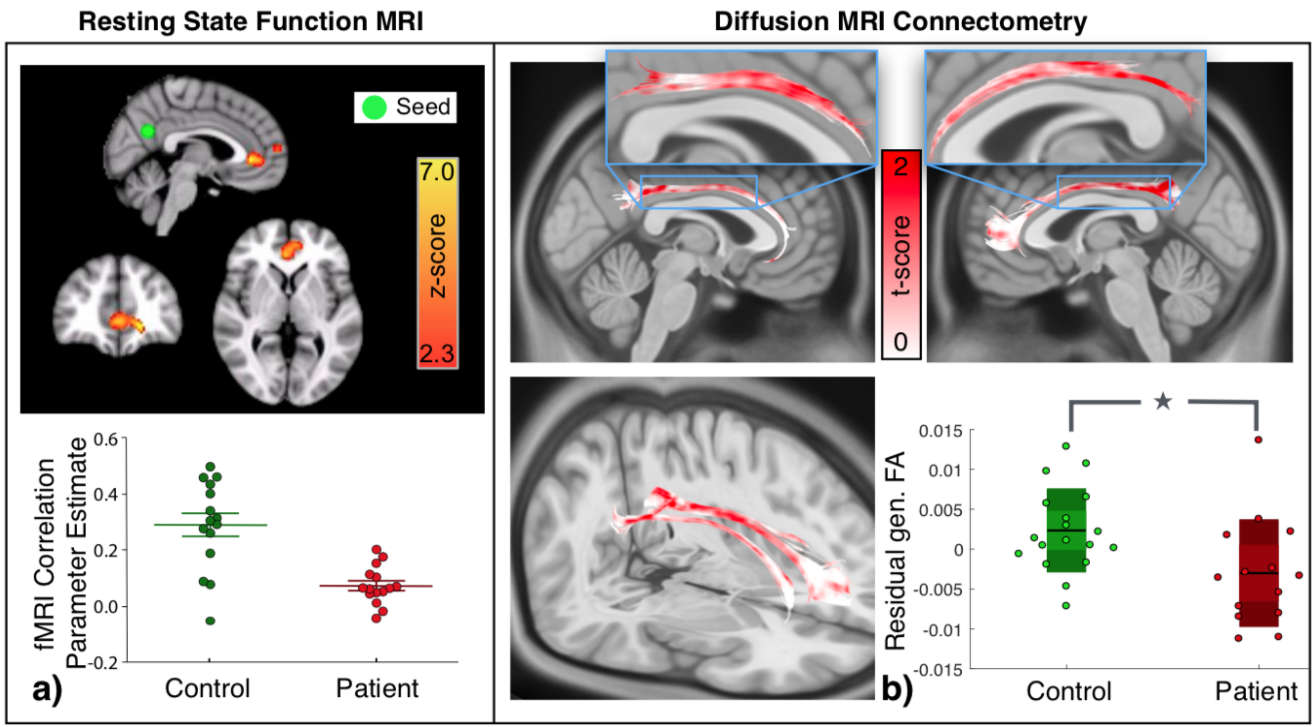

Figure 2: **Microstructural white matter alterations in default mode network detected from connectometry analysis correspond to functional alterations. a)** Functional connectivity between posterior cingulate cortex (PCC) and ventral part of medial prefrontal cortex (MPFC) has been shown to be reduced in patients with IGE (figure modified from McGill et al. 2012). **b)** The underlying cingulum tracts between PCC and MPFC obtained from tractography between these regions are shown. Implementing Connectometry analysis elucidates regional differences in gFA between the two groups along the tract profile which are colour coded in accordance with their *t*-score. Red represents the tract segments that are maximally reduced in patients whereas white illustrates the segments that are not different. The specific portion of cingulum tracts which are aberrant are magnified in the inset figures. The box plot depicts that the residuals of mean gFA associated with the aforementioned aberrant portion of cingulum tracts is significantly reduced ($p = 0.01$, $d = 0.90$) in patients compared to controls.

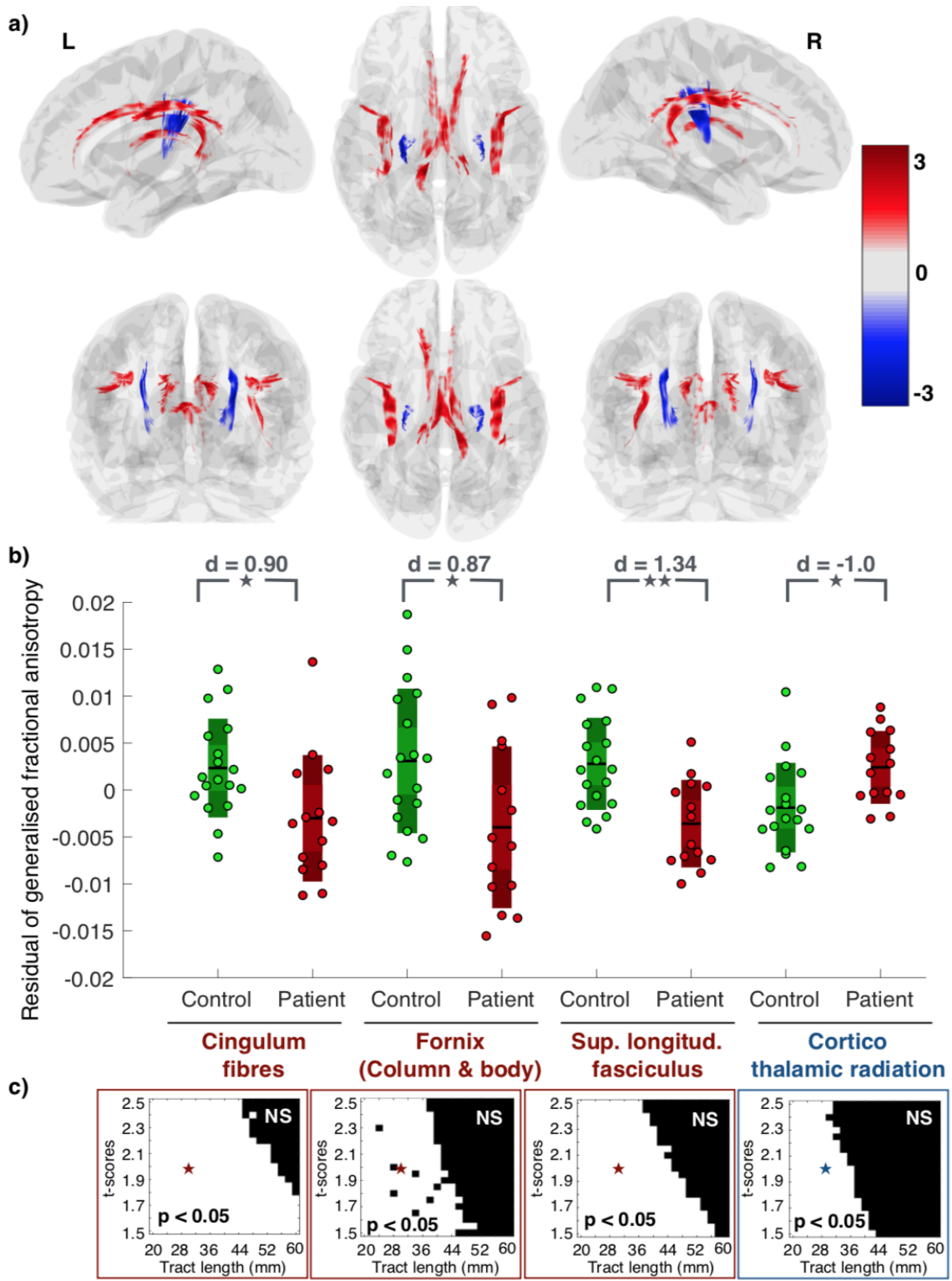

Figure 3: **Alterations in generalised fractional anisotropy in white matter tracts. a)** gFA associated with the white matter tracts in bilateral cingulum, column and body of fornix, superior longitudinal fasciculus, and the parts of cortico-thalamic radiations terminating in the precentral gyrus are aberrant for a wide range of thresholds in connectometry analysis. These tracts are shown overlaid on the brain schematic in different views. Tracts are colour coded in accordance with their *t*-scores where the warmer colour for higher positive *t*-score indicates a reduced gFA in patients and the cooler colour for lower negative *t*-score indicates increased gFA associated with the tracts in patients compared to controls. **b)** The comparison of mean gFA across the tracts between patients and controls for each aberrant region is shown in the box plot. The *p*-value for each region is indicated by stars in grey (double star for $p < 0.005$ and single star for $0.005 < p < 0.05$) along with the effect sizes (*d*-scores). **c)** The white region in the two dimensional panels depict the threshold values for which the micro-structural alterations are significantly different in connectometry analysis. Stars in red/blue represent exemplary threshold values of $t = 1.94$ and tract length = $30mm$ at which the results are shown in **a)** and **b)**.

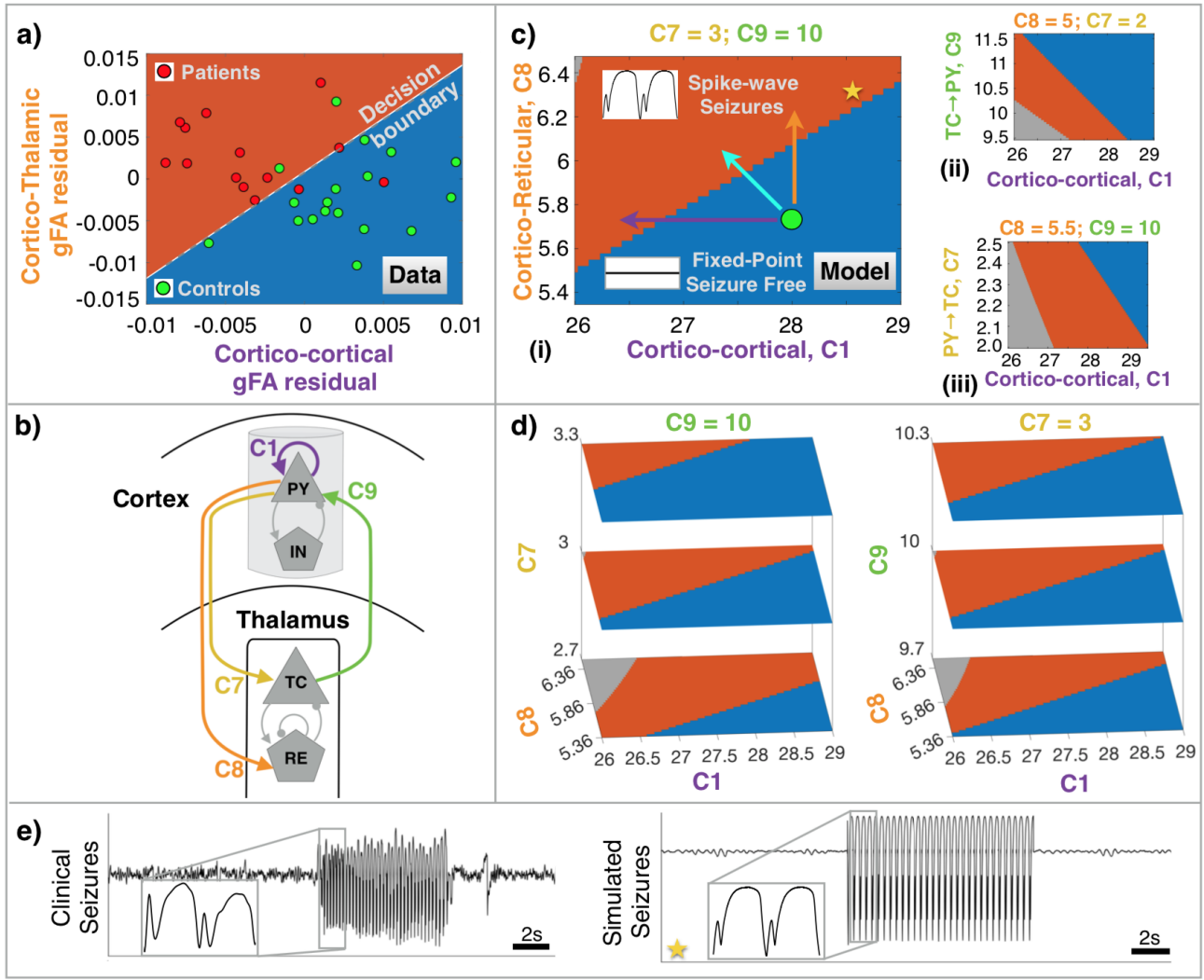

Figure 4: **Computational model combined with neuroimaging analysis predicts potential mechanism of epilepsy manifestation.** **a)** Discriminatory decision boundary between patients and controls inferred from a logistic regression model by incorporating the residuals of mean gFA of combined cortico-cortical and cortico-thalamic alterations as features. **b)** Schematic of thalamo-cortical dynamical model. C1 represents the lumped cortico-cortical connections, C9 is the connection from thalamus to cortex, C7 indicates the connection from cortex to thalamocortical relay cells, and C8 is the connection from cortex to thalamic reticular nucleus. **c)** Model dynamics colour-coded by behaviour in parameter space. C1 is shown against C7, C8, and C9, with representative two-dimensional slices of the parameter space. The blue region marks the fixed-point dynamics representative of healthy seizure-free activity. The orange and grey regions exhibit spike-wave and fast oscillatory seizure dynamics. Arrows indicate potential ways in altering connectivity that can render a healthy brain to become epileptogenic, according to the model. **d)** Three dimensional slices indicating that the bifurcation orientation is preserved in C8-C1 slices even with changes in C7 and C9 parameters. **e)** An example clinical seizure in IGE is compared with the one simulated from the model in the bistable regime. The model parameter is marked with a corresponding star in **c)**.

24